\documentclass[10pt,letterpaper,twocolumn]{article} 

\usepackage{ol2}
\usepackage{hyperref}
\usepackage{amsmath}
\usepackage{graphicx}

\begin{document}

\twocolumn[ 

\title{Extending dark optical trapping geometries}


\author{Aidan S. Arnold$^*$}

\address{Department of Physics, SUPA, University of Strathclyde, Glasgow G4 0NG, UK\\
$^*$Corresponding author: aidan.arnold@strath.ac.uk }

\begin{abstract}
New counter-propagating geometries are presented for localising ultracold atoms in the dark regions created by the interference of Laguerre-Gaussian laser beams. In particular dark helices, an `optical revolver,' axial lattices of rings and axial lattices of ring lattices of rings are considered and a realistic scheme for achieving phase-stability is explored. The dark nature of these traps will enable their use as versatile tools for low-decoherence atom interferometry with zero differential light shifts.
\end{abstract}


 ] 

The optical dipole force generated by far off-resonance laser beams has been utilised in an extremely wide variety of experiments on ultracold matter. It has been used as a matter wave beamsplitter/tunable barrier \cite{plug1,plug2}, has stirred up vortices \cite{vort}, and created crystals of atoms stored in precise optical lattices \cite{lattice}.
If the light is red-detuned from an atomic resonance atoms are attracted to regions of bright optical intensity, whereas blue-detuned light attracts atoms to the lowest intensity regions.   

Smoothly varying electric fields can have a zero crossing in space, and one can thereby make spatial regions in one, two or three dimensions which are completely dark. The advantage of this darkness is that as the atoms scatter photons at a rate proportional to $I/\Delta^2,$ where $I$ is the light intensity and $\Delta$ is the detuning from resonance, ultracold atoms in dark traps have a greatly reduced heating rate due to photon absorption and emission. 

In addition, the light shift due to the optical dipole effect is proportional to $I/\Delta$, and thus two regions of different intensity $I_1$ and $I_2$ will have corresponding energy difference $E_2-E_1$, leading to an (often unwanted) time-dependent phase shift $\phi=(E_2-E_1)t/\hbar.$ Such a problem is obviated in dark traps. Gravitational forces and thermal motion/quantum mechanical zero-point energy will lead to sampling of non-dark regions of the potential, however magnetic `levitation' \cite{plug2} and ultralow-temperatures, respectively, can minimise these effects.

To illustrate the wide range of options available with optical trapping, we discuss some of the geometries for atom trapping obtained to date. Single-beam blue-detuned traps can be used as `optical pipes' \cite{bongs} via Laguerre-Gauss (LG or `donut') laser modes \cite{sonjarev}. By combining two co-propagating laser beams one can also make 3D dark traps `optical bottles' \cite{davidson,miles}, rings \cite{fatemi} and ring lattices (`optical ferris wheels') \cite{ferris,zhan}. To date \textit{counter}-propagating geometries with LG beams have been relatively rarely used experimentally, although they open a plethora of geometries as we show here. Cylindrical axicon beams without the angular momentum of LG beams were used in a counter-propagating geometry in Ref.~\cite{hennequin} to create a dark axial array of rings (`axial' refers to the beam propagation direction). Theoretical papers have discussed ways to make bright (red-detuned) axial lattices of ring lattices \cite{amico}, and helices \cite{bhatta}, and we extend all of these geometries into the preferable (low decoherence and light shift) dark configuration.   


The Laguerre-Gaussian field mode of index $\ell$ (only the case with the other index $p=0$ is used) can be expressed in cylindrical polars as:
\begin{eqnarray}
LG_{\ell}\!\! & = & \!\!\sqrt{\frac{2^{1+|\ell|}P \, r^{2 |\ell|}}{\pi w^{2(1+|\ell|)} |\ell|!}}\; 
\exp\!\left[i s k z \left(1+\frac{r^2}{2 (z^2+{z_R}^2)}\right)\right] \nonumber \\
 & & \!\! e^{-r^2/w^2-i\ell\theta-i \omega t-i s (|\ell|+1) \arctan(z/z_R)} 
\label{mode}
\end{eqnarray}
where the beam has wavenumber $k=2\pi/\lambda,$ angular frequency $\omega=c k,$ power $P,$ waist $w_0$, Rayleigh range $z_R=\pi {w_0}^2/\lambda,$ $w(z)=w_0\sqrt{1+z^2/{z_R}^2}$ and $s=\pm1$ is the propagation direction relative to the $z$ axis. In the following illustrations we will consider rubidium atoms in light fields of wavelength $\lambda=532\,$nm, far blue-detuned from the atomic resonances at $\lambda_0=780\,$nm and $795\,$nm. All beam waists are $w_0=5\,\mu$m, unless stated otherwise. Note the mode described in Eq.~\ref{mode} is the square-root of the intensity (i.e.\ not the electric field, but proportional to it). Useful estimates of the scattering rate and potential depth of a dipole beam light with intensity $I$ are \cite{ferris}:
 \[(R,U)\approx \hbar \Gamma \beta \left({\Delta_\Gamma}^{-2}, (k_B \Delta_\Gamma)^{-1} \right)/8 \] respectively, where $\beta=I/I_S$ and the Rb saturation intensity and natural linewidth are $I_S=16.3$W/cm$^2$ and $\Gamma=2\pi\times6\,$MHz, respectively. The detuning of the dipole beam in linewidths is $\Delta_\Gamma=(\omega-\omega_0)/\Gamma\approx 3.0\times 10^7.$ 

\begin{figure}[!b]
\begin{minipage}{.47\columnwidth}
\includegraphics[width=.975\columnwidth]{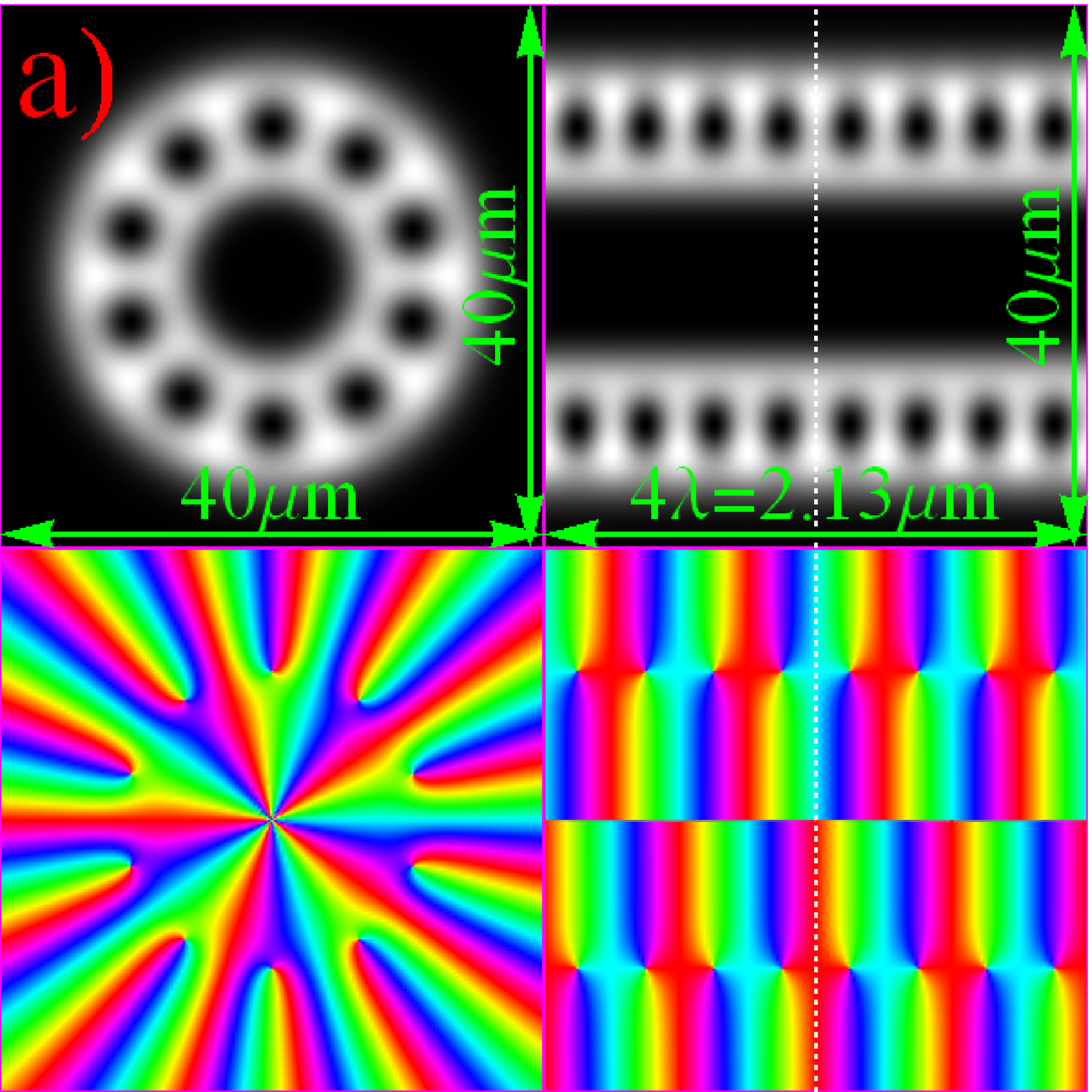}
\end{minipage}
\begin{minipage}{.47\columnwidth}
\includegraphics[width=\columnwidth]{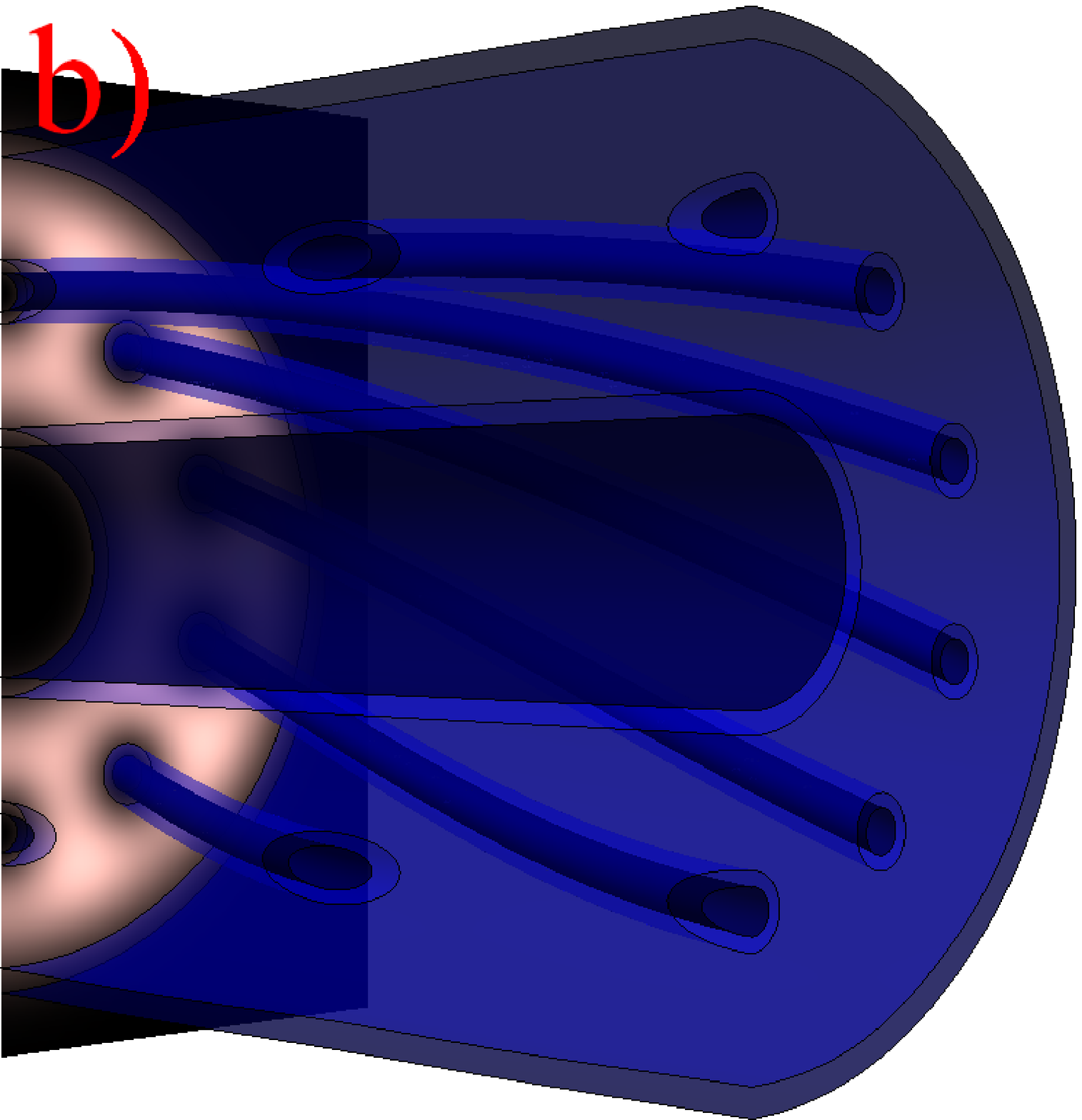}
\end{minipage}
 \caption{A dark helix lattice arising from a superposition of $100\,$mW of $LG_5$ and a counterpropagating $170\,$mW $LG_{15}$ beam. Image a) contains radial (left $(40\,\mu$m$)^2$)) and axial (right $(4\lambda=2.13\,\mu$m$)\times 40\,\mu$m) slices through the beam. Intensity and phase are shown in the top and bottom images, respectively. At the maximum intensity shown (white zones) the scattering rate and potential depth are $R=0.19\,$Hz and $U=43\,\mu$K, respectively. Image b) extends the radial 2D intensity plot into a 3D contour plot indicating the intensity minima (blue) at 10\% and 25\% of the maximum of the scale in the radial intensity slice. Note the axial extension is only one wavelength $532\,$nm, i.e.\ exaggerated by a factor of $\approx 75$.} \label{fighelix}
\end{figure}

\begin{figure}[!b]
\begin{minipage}{.47\columnwidth}
\includegraphics[width=.985\columnwidth]{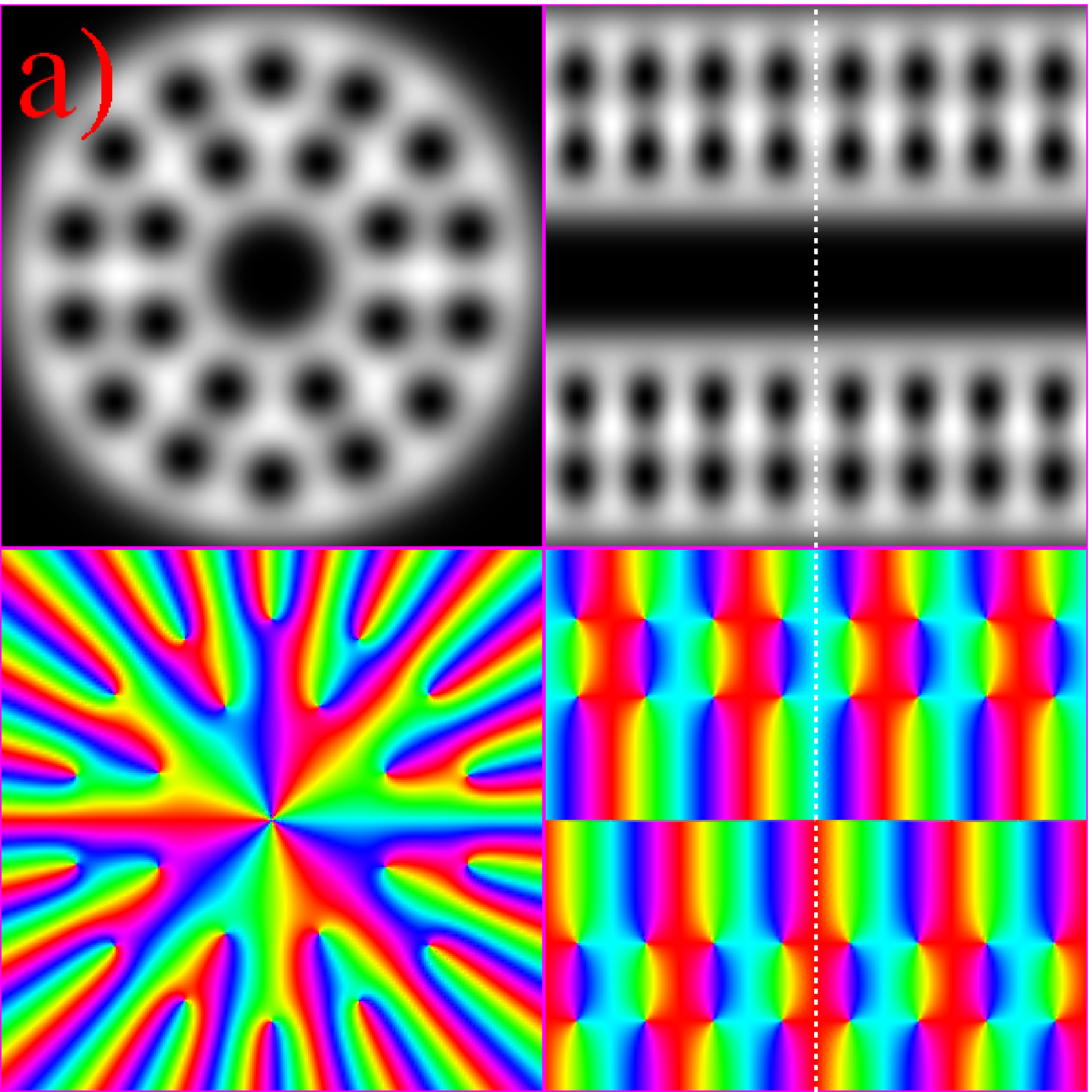}
\end{minipage}
\begin{minipage}{.47\columnwidth}
\includegraphics[width=\columnwidth]{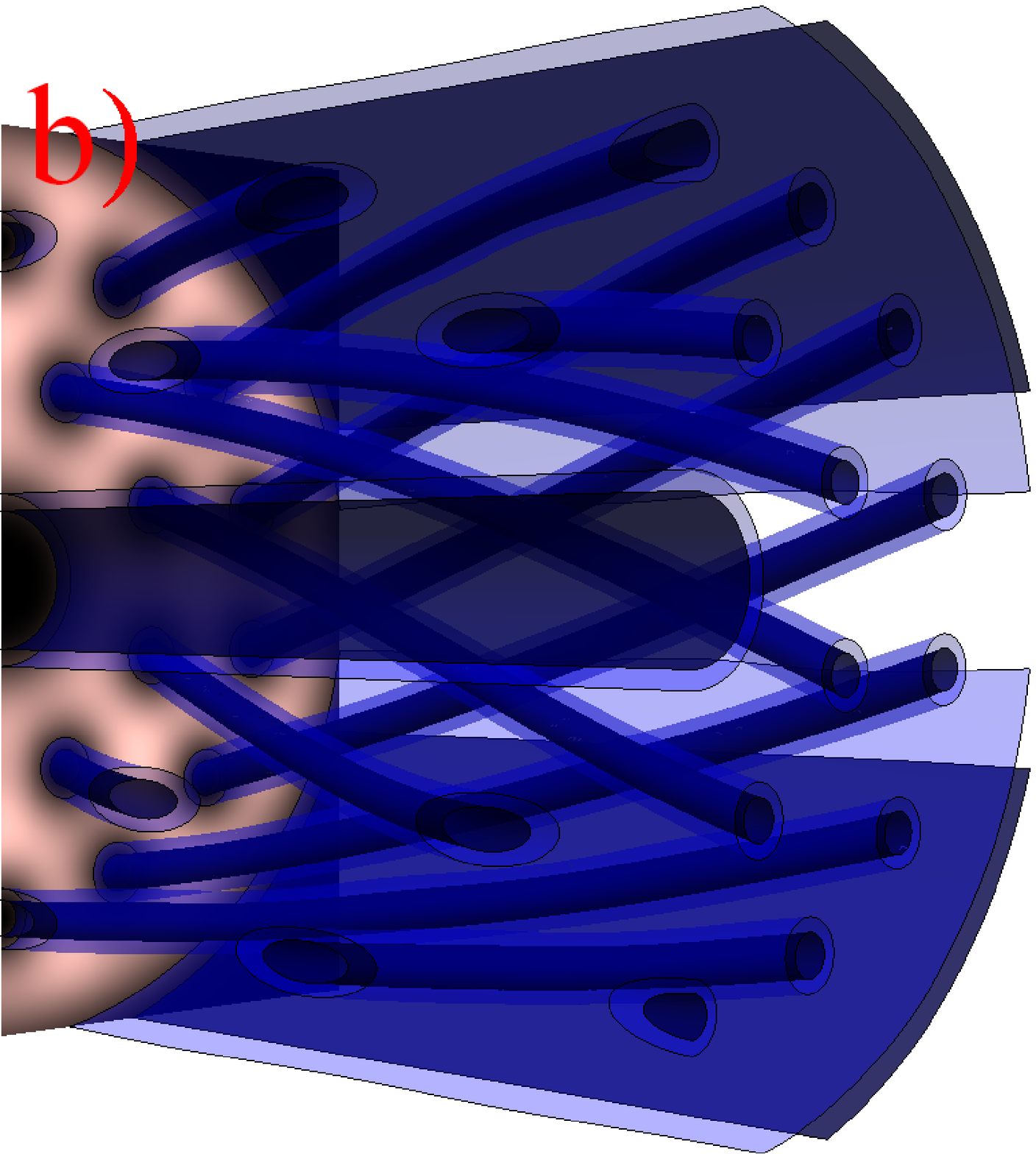}
\end{minipage}
 \caption{An `optical revolver' generated by a superposition of $100\,$mW of $LG_3$ and $290\,$mW of  $LG_{25}$ with counterpropagating $190\,$mW of $LG_{11}$, all other dimensions as for Fig.~\ref{fighelix}. Here $R=0.26\,$Hz, $U=60\,\mu$K. The inner helical dark core has a lead angle of $38\,$mrad and leads to harmonic trap frequencies $183\,$kHz and $6.9\,$kHz in the `axial' and radial directions, respectively.} \label{figrevolve}
\end{figure}

\begin{figure}[!b]
\begin{minipage}{.47\columnwidth}
\includegraphics[width=.985\columnwidth]{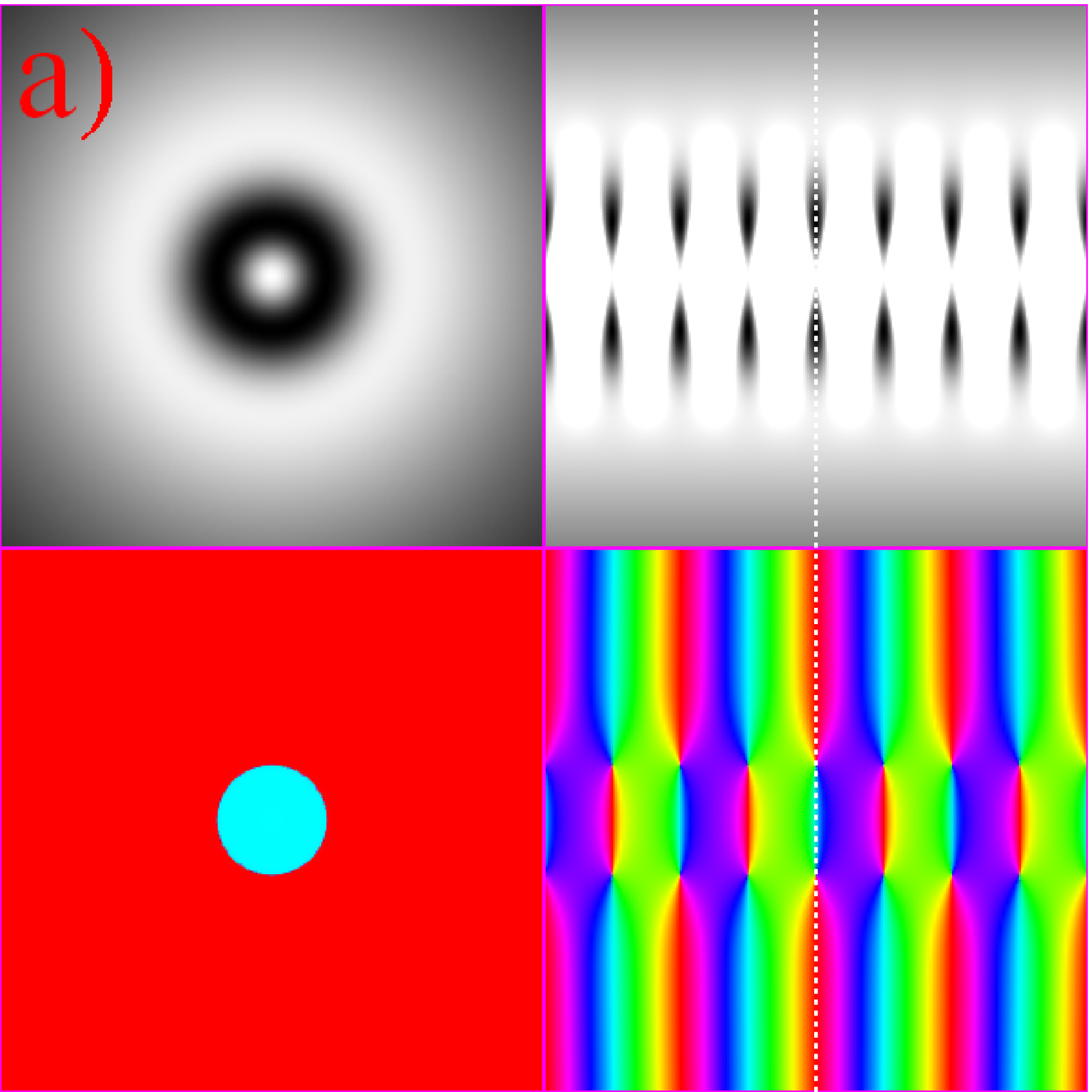}
\end{minipage}
\begin{minipage}{.47\columnwidth}
\includegraphics[width=\columnwidth]{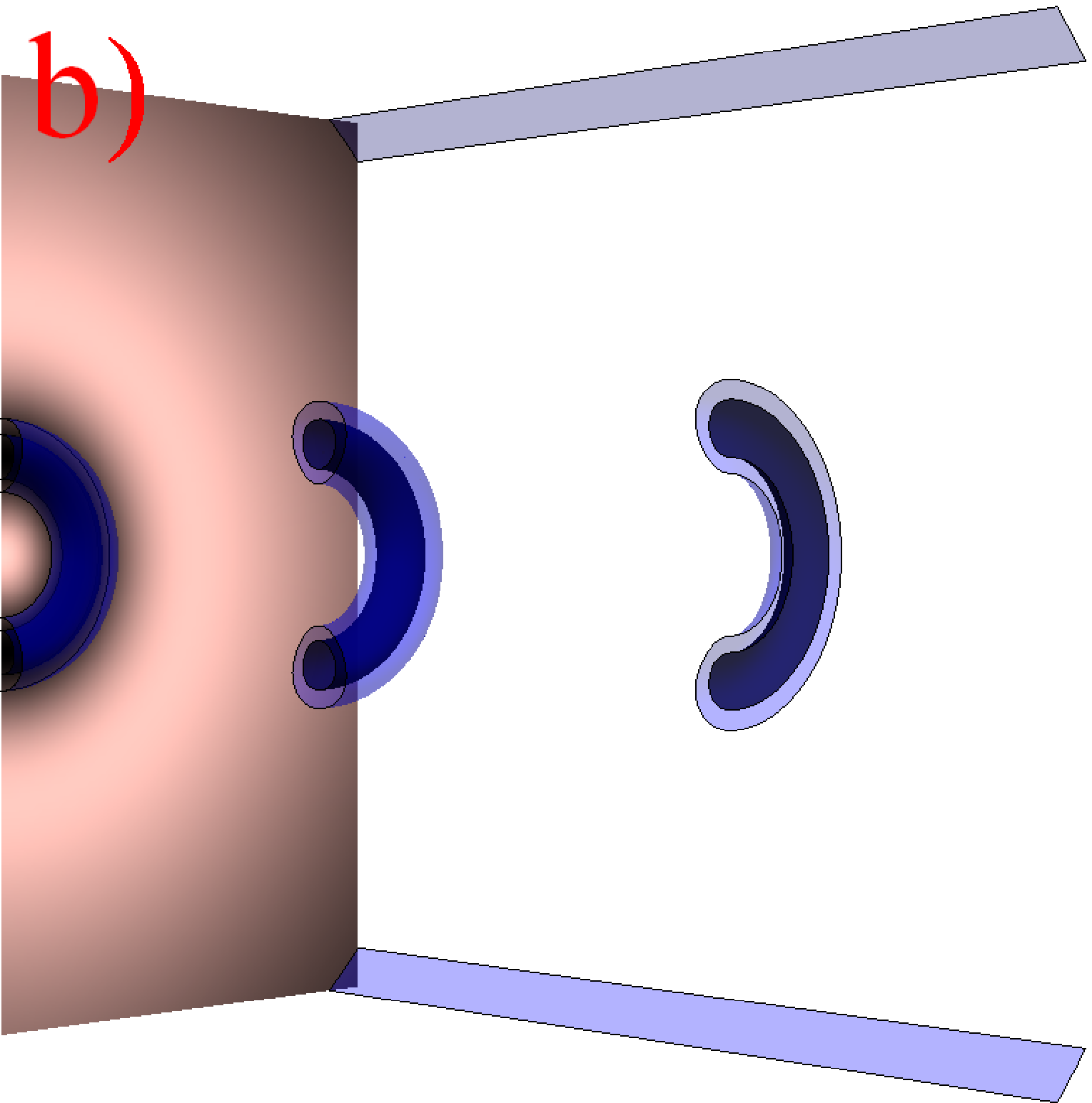}
\end{minipage}
\vspace{-1mm} \caption{An axial lattice of rings formed by a $100\,$mW $LG_0$ beam (with the typical $5\,\mu$m waist) counterpropagating with $1.0\,$W of $LG_0$ beam (with a waist of $30\,\mu$m). Here $R=0.18\,$Hz and $U=42\,\mu$K.  Each dark ring has trap frequencies $370\,$kHz and $4.9\,$kHz in the axial and radial directions, respectively. This geometry is useful for storing multiple persistent currents or creating annular Tonks-Girardeau gases.} \label{figLR}
\end{figure}

\begin{figure}[!b]
\begin{minipage}{.47\columnwidth}
\includegraphics[width=.985\columnwidth]{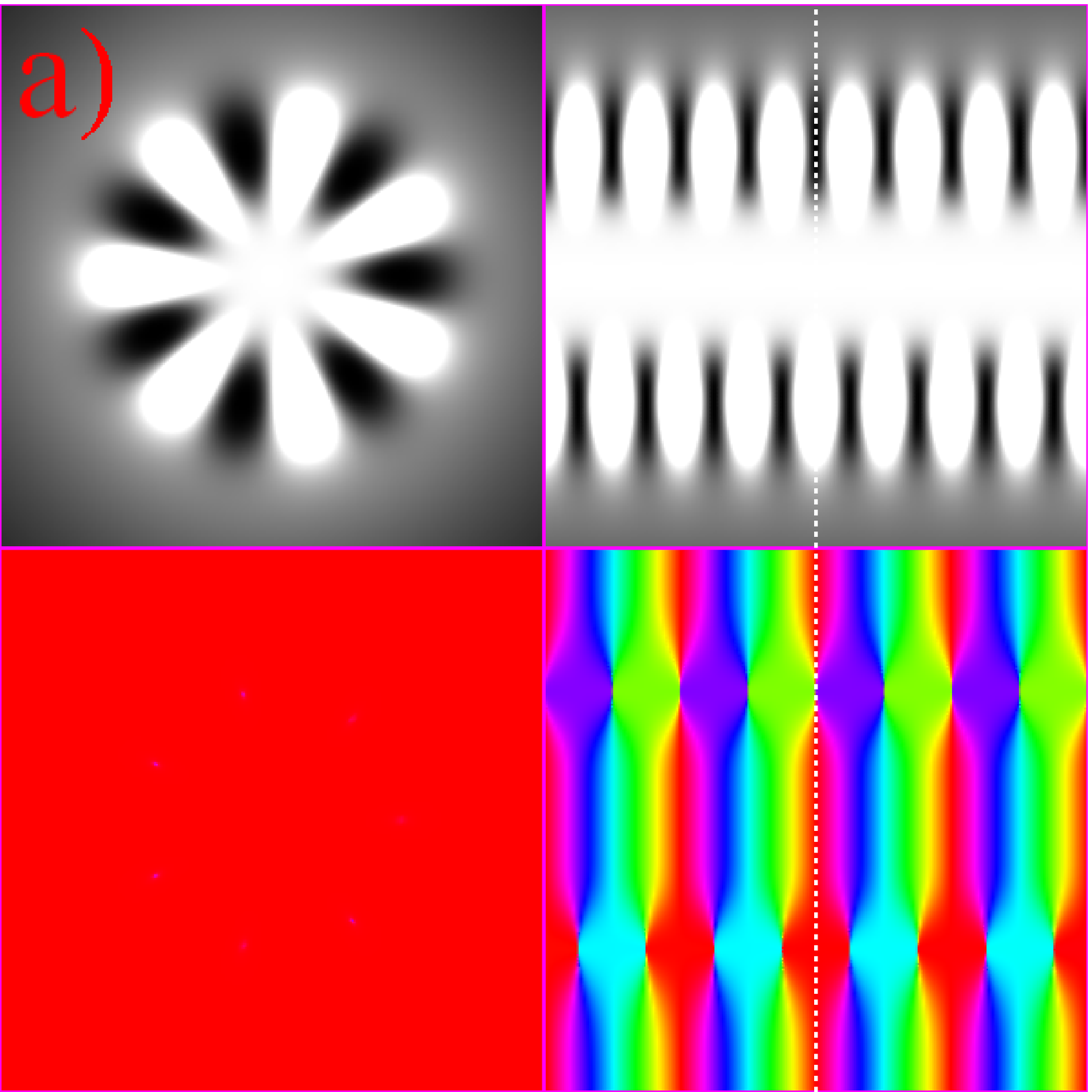}
\end{minipage}
\begin{minipage}{.47\columnwidth}
\includegraphics[width=\columnwidth]{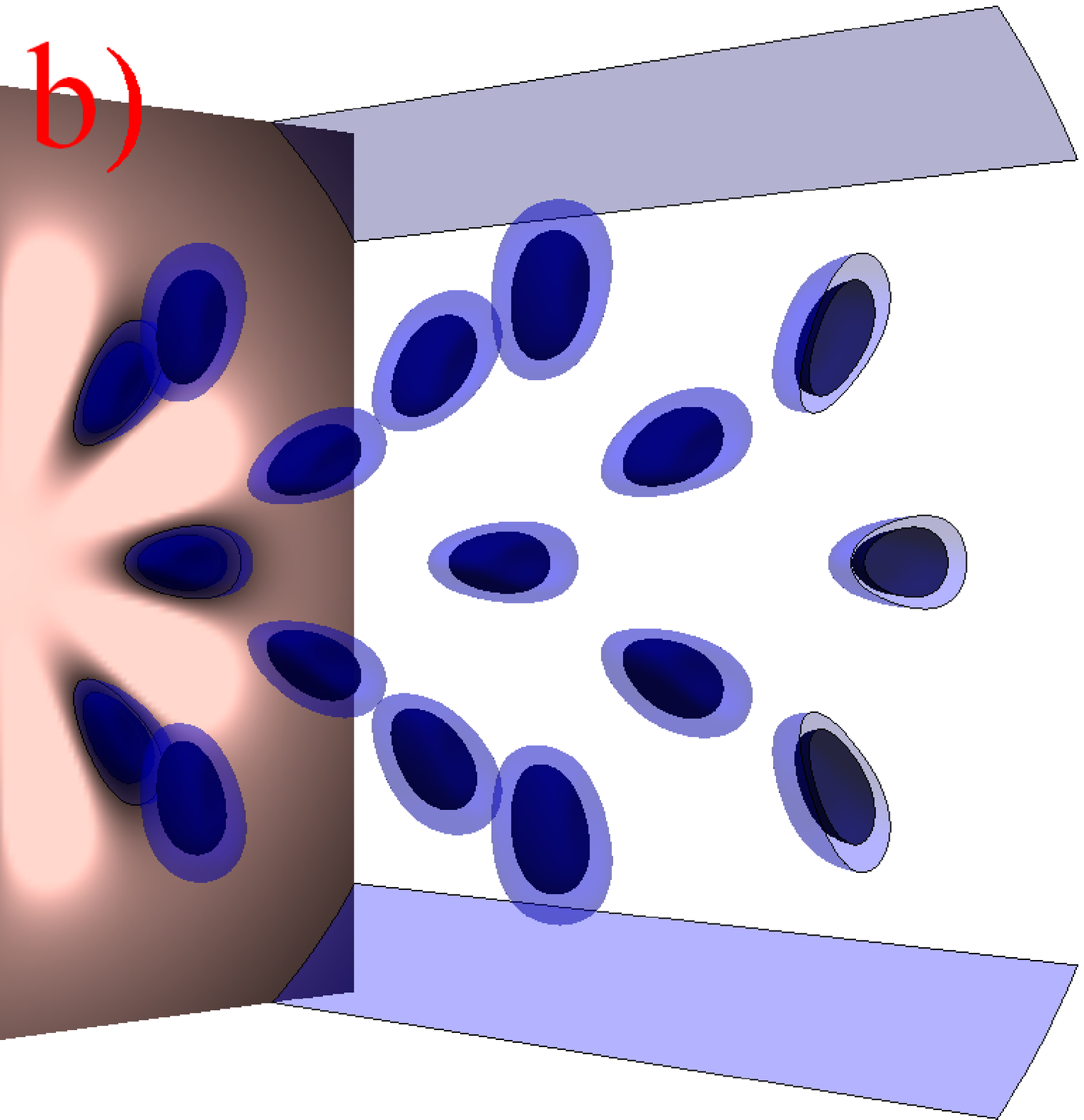}
\end{minipage}
\vspace{-1mm} \caption{An axial lattice of ring lattices formed by a $100\,$mW $LG_7$ beam and $100\,$mW $LG_{-7}$ (with the typical $5\,\mu$m waist) counterpropagating with $2.6\,$W of $LG_0$ beam (with a waist of $30\,\mu$m). Here $R=0.60\,$Hz and $U=140\,\mu$K. Each dark trap has an axial trap frequency of $550\,$kHz, and the trap is quartic rather than harmonic in the radial and azimuthal directions -- useful for exploring 2D quantum gases.} \label{figLRLR}
\end{figure}

\begin{figure}[!b]
\begin{minipage}{.495\columnwidth}
\includegraphics[width=.985\columnwidth]{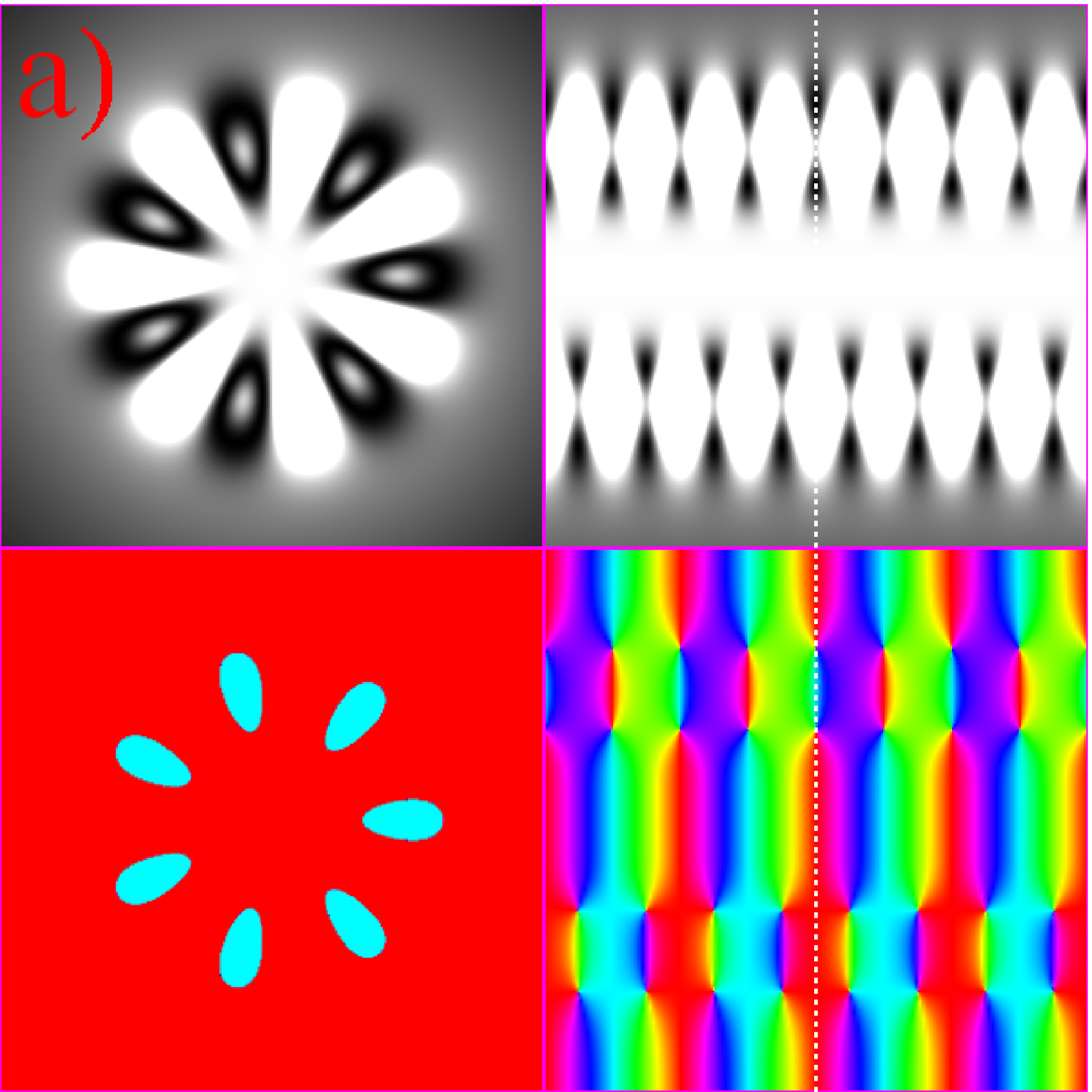}
\end{minipage}
\begin{minipage}{.495\columnwidth}
\includegraphics[width=\columnwidth]{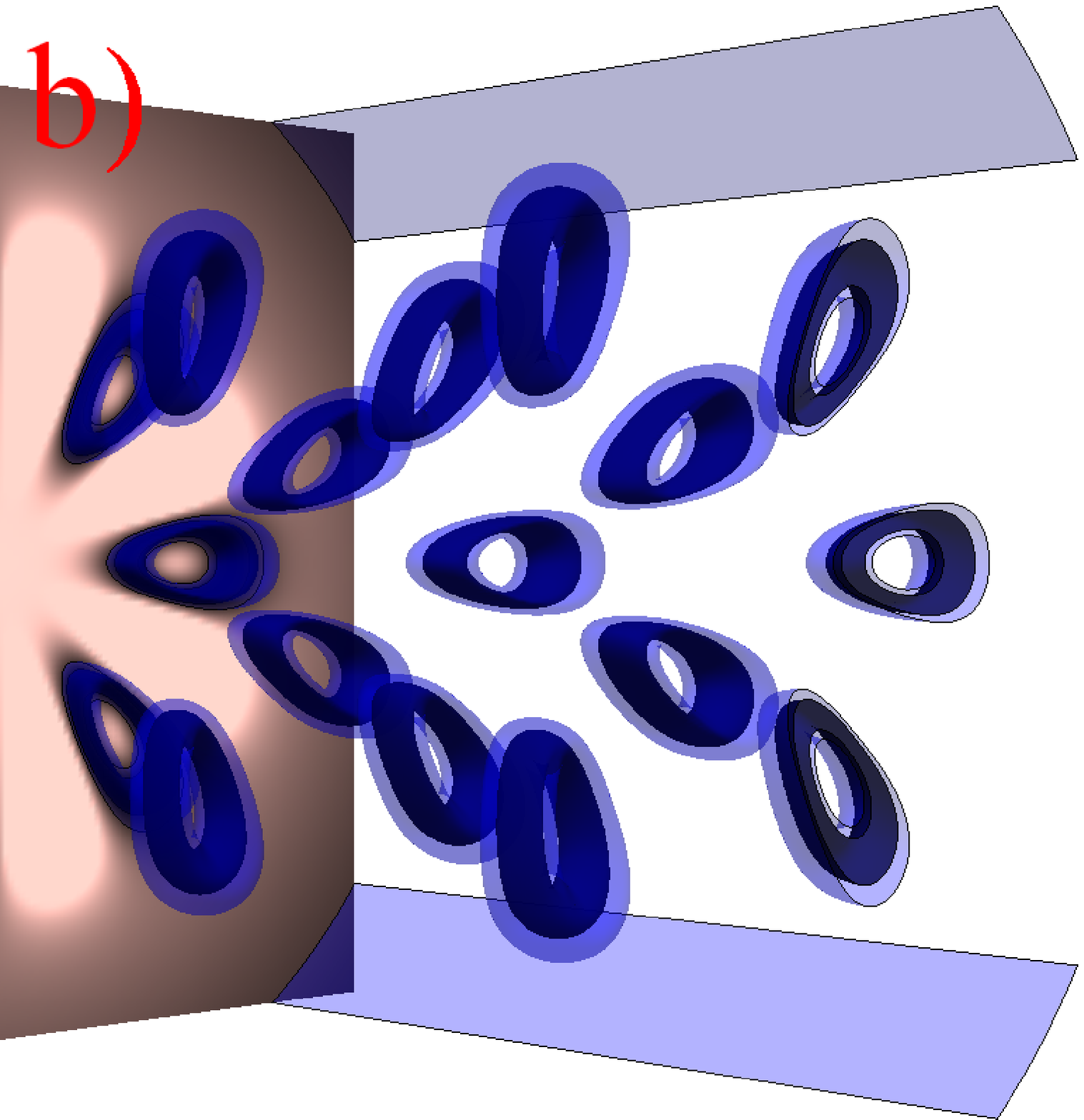}
\end{minipage}
\vspace{-1mm} \caption{An axial lattice of ring lattices of rings, all parameters as in Fig.~\ref{figLRLR}, except the counterpropagating Gaussian beam has power $650\,$mW. Here $R=0.15\,$Hz and $U=34\,\mu$K.  Each dark ring has a spatially varying axial (radial) trap frequency of $260-290\,$kHz ($4.6-15\,$kHz).  
} \label{figLRLRR}
\end{figure}

Examples of the different kinds of dark optical trap geometries obtainable using superpositions of LG beams are: dark helix lattices (Fig.~\ref{fighelix}), counter-rotating dark helix lattices (Fig.~\ref{figrevolve}), an axial lattice of rings (Fig.~\ref{figLR}), an axial lattice of ring lattices (Fig.~\ref{figLRLR}) and an axial lattice of ring lattices of rings (Fig.~\ref{figLRLRR}). 

Tonks-Girardeau gases are intriguing systems where 1D confinement of strongly-interacting bosons can lead to behavior mimicking non-interacting fermions \cite{tonks}. In Fig.\ \ref{fighelix} (similar to Fig.\ \ref{figrevolve}) each helical dark core has a lead angle $\approx\frac{\lambda(\ell_1-\ell_2)}{\sqrt{2}\pi w_0 (\sqrt{\ell_1}+\sqrt{\ell_2})}=39\,$mrad and has harmonic trap frequencies $154\,$kHz and $6.0\,$kHz in the `axial' and radial directions, respectively. By reducing the detuning (to tighten the trap) it will be possible to use each helix to compactly store long Tonks gas chains -- in Fig.\ \ref{fighelix} even a $40 \mu$m axial extent of helix with radius $11\,\mu$m has an uncoiled length of $1\,$mm. Such an extension is not possible in crossed-beam dipole lattices where trap lengths are limited to the typical beam waists of $\approx 100\,\mu$m. For the helical scheme tunable axial confinement would be provided with harmonic magnetic fields.

The use of larger rings for trapping Bose-Einstein condensates has applications in rotational sensing \cite{aa}, and more recently there have been rapid developments in small-scale optical traps for precise studies of superfluidity and the quantization of flow in rings \cite{phillips,zoran}. The traps we show in Figs.\ \ref{figLR} and \ref{figLRLRR} illustrate a way of extending the single ring traps of Refs. \cite{phillips,zoran}, allowing scalability to a large array of rings in which the flows in each ring could be stored as individual qubits for quantum processing. 

A major advantage of the counter-propagating lattice geometry is that structures can be formed with sub-wavelength axial structure, and tight axial confinement can be achieved without additional fields -- like the bright geometry discussed in Ref.\ \cite{amico}. The lattice is an interferometric superposition of counter-propagating beams, as such although the shape of the lattice is not phase-dependent, phase-stability is required to completely axially localize the optical traps. A trick used in optical lattices, to avoid aligning two separate beams with interferometric stability, is to retroreflect Gaussian beams with a mirror near the position of the trap. The node of the transverse electric field at the mirror ensures lattice stability if this mirror alone is stable. 

\begin{figure}[!b]
\includegraphics[width=.985\columnwidth]{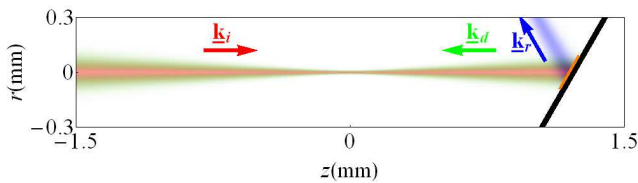}
\vspace{-3mm} \caption{An incident (red, \underline{\textbf{k}}$_i$) laser beam diffracts from a 2D phase grating (with a lens) into counter-propagating (green, \underline{\textbf{k}}$_d$) and discarded reflected (blue, \underline{\textbf{k}}$_r$) orders.} \label{scheme}
\end{figure}

Here we suggest (Fig.~\ref{scheme}) the use of a phase grating near the trap location as this will have three advantages: the lattice axial location depends only on the grating stability, alignment of the trap relies only on successful angle alignment of the grating, and the grating itself can be used to change the LG mode superposition of the beam into the required counter-propagating mode to complete the lattice. In order to achieve sufficient resolution that the grating efficiently and accurately generates the required mode, the incoming beam must cover several hundred grating periods. This could be achieved in vacuo, in Littrow configuration, slightly away from the beam focus with a micro-fabricated grating substrate with a size of $\approx 200\,\mu$m and grating period $\approx1\,\mu$m (ensuring a large diffraction angle so other diffracted orders do not affect the beam superposition). An ex-vacuo grating would simplify alignment, and as long as an astigmatism-compensated zoom lens is used \cite{kuhr}, a spatial light modulator could be used as a means to simply program different phase gratings and change lattice geometries.

The scheme's robustness to experimental imperfections is vital. Laser pointing stability is paramount, precluding ion lasers in favour of fiber-coupled tapered amplifiers or doubled DPSS/fiber lasers.  Counter-propagating beam misalignment can be minimised using a small aperture near the fiber output, or by fiber-recoupling counter-propagating light. We estimate this misalignment, via ray-tracing a simple beamline, to be angular (and associated radial) values of $\approx 1\,$mrad and $1\,\mu$m, at the trap location. This leads to noticeable  changes in the total intensity pattern, particularly for Figs.~\ref{fighelix}a and \ref{figrevolve}a due to the small counter-propagating waists. However, the dark trapping potentials in the 3D contour plots (Figs.~\ref{fighelix}-\ref{figLRLRR} b) remain largely unchanged, even with relative mode power changes of $\approx 20\%$.
In conclusion we have illustrated several new dark counter-propagating optical trapping geometries and presented a realistic, robust scheme for their experimental generation. The dark nature of these traps mean that their zero differential light shifts can be used in atom interferometry, and they also have applications in extending Tonks-Girardeau gases, parallelisation of superflow and quantum information storage.

I am grateful for stimulating discussions with E.\ Riis, P.\ F.\ Griffin, S.\ Franke-Arnold and G.\ Walker.

\bibliographystyle{osajnl}

\end{document}